\documentclass[aps,prb,twocolumn,showpacs,showkeys,footinbib,superscriptaddress]{revtex4-1}%superscriptaddress

\usepackage{amsmath}
\usepackage{amssymb}
\usepackage{graphicx}
\usepackage{bm}
\usepackage{color}
\usepackage{relsize}
\usepackage{braket}

\usepackage{hyperref}
\usepackage[all]{hypcap}

\renewcommand{\i}{\ensuremath{\mathrm{i}}}
\newcommand{\e}{\ensuremath{\mathrm{e}}}
\renewcommand{\d}{\ensuremath{\mathrm{d}}}

\newcommand{\diag}{\mathrm{diag}}

\begin{document}
\title{Probing topological transitions in HgTe/CdTe quantum wells by magneto-optical measurements}
\author{Benedikt Scharf}
\affiliation{Department of Physics, University at Buffalo, State University of New York, Buffalo, NY 14260, USA}
\affiliation{Institute for Theoretical Physics, University of Regensburg, 93040 Regensburg, Germany}
\author{Alex Matos-Abiague}
\affiliation{Department of Physics, University at Buffalo, State University of New York, Buffalo, NY 14260, USA}
\author{Igor \v{Z}uti\'c}
\affiliation{Department of Physics, University at Buffalo, State University of New York, Buffalo, NY 14260, USA}
\author{Jaroslav Fabian}
\affiliation{Institute for Theoretical Physics, University of Regensburg, 93040 Regensburg, Germany}

\date{\today}

\begin{abstract}
In two-dimensional topological insulators, such as inverted HgTe/CdTe quantum wells, helical quantum spin Hall (QSH) states persist even at finite magnetic fields below a critical magnetic field $B_\mathrm{c}$, above which only quantum Hall (QH) states can be found. Using linear-response theory, we theoretically investigate the magneto-optical properties of inverted HgTe/CdTe quantum wells, both for infinite two-dimensional and finite-strip geometries, and for possible signatures of the transition between the QSH and QH regimes. In the absorption spectrum, several peaks arise due to non-equidistant Landau levels in both regimes. However, in the QSH regime, we find an additional absorption peak at low energies in the finite-strip geometry. This peak arises due to the presence of edge states in this geometry and persists for any Fermi level in the QSH regime, while in the QH regime the peak vanishes if the Fermi level is situated in the bulk gap. Thus, by sweeping the gate voltage, it is possible to experimentally distinguish between the QSH and QH regimes due to this signature. Moreover, we investigate the effect of spin-orbit coupling and finite temperature on this measurement scheme.
\end{abstract}


\maketitle

\section{Introduction}\label{Sec:Intro}

\begin{figure}[t]
\centering
\includegraphics*[width=8.65cm]{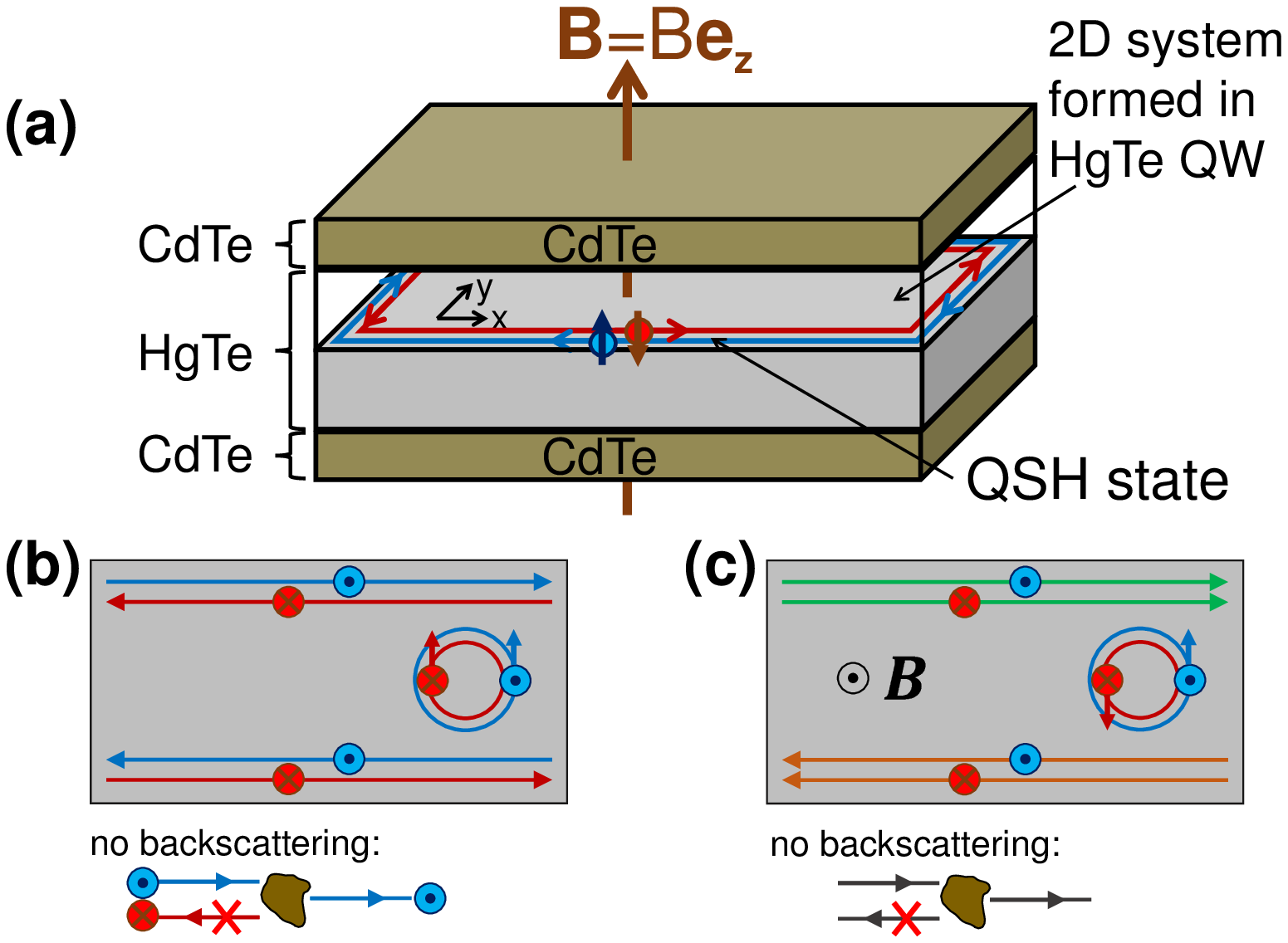}
\caption{(Color online) Schematic views of the (a) HgTe/CdTe QWs considered in this work as well as of the (b) QSH and (c) QH edge states and classical bulk orbits. Here, dotted and crossed circles denote spin-up and spin-down electrons, respectively, while arrows denote their respective directions of motion.}\label{fig:Scheme}
\end{figure}

Since the first theoretical predictions of the quantum spin Hall (QSH) effect in graphene\cite{Kane2005:PRL,Kane2005:PRL2} and in inverted HgTe/CdTe quantum-well (QW) structures,\cite{Bernevig2006:Science} topological insulators and topological superconductors have evolved into a topic of immense research interest in recent years.\cite{Hasan2010:RMP,Qi2011:RMP,Shen2012,Culcer2012:PhysicaE,Tkachov2013:PSS} Shortly after those proposals, the QSH state has first been demonstrated experimentally in inverted HgTe/CdTe QWs [see Fig.~\ref{fig:Scheme}~(a)],\cite{Koenig2007:Science,Koenig2008:JPSJ,Buettner2011:NatPhys,Bruene2012:NatPhys} where one can tune the band structure by fabricating QWs with different thicknesses $d$.\cite{Liu2008:PRL} Furthermore, several other two-dimensional (2D) systems, such as GaAs under shear strain,\cite{Bernevig2006:PRL} 2D bismuth,\cite{Murakami2006:PRL} or inverted InAs/GaSb/AlSb semiconductor QWs with type-II band alignment,\cite{Liu2008:PRL} have been proposed theoretically to exhibit QSH states, and three-dimensional analogs of the QSH states have also been found, both theoretically\cite{Fu2007:PRL} and experimentally,\cite{Hsieh2008:Nature,Hsieh2009:PRL} giving rise to the concept of topological insulators.\cite{Xu2006:PRB,Moore2007:PRB}

Topological insulators are materials that are insulating in the bulk, but possess dissipationless edge or surface states, whose spin orientation is determined by the direction of the electron momentum. In analogy to the helicity, which describes the correlation between the spin and the momentum of a particle, those spin-momentum locked edge or surface states have been termed helical. A 2D topological insulator, synonymously also referred to as a QSH insulator, and its helical edge states are illustrated in Fig.~\ref{fig:Scheme}~(b): Due to their helical nature, those topological edge states are counterpropagating, spin-polarized states, which are protected against time-reversal invariant perturbations such as scattering by nonmagnetic impurities, and thus also of interest for spintronics applications.\cite{Zutic2004:RMP,Fabian2007:APS,Mahfouzi2012:PRL,Zhang2014:PRL,Triola2014:PRB} These QSH states are in sharp contrast to quantum Hall (QH) states, which emerge if a magnetic field is applied and which are chiral in the sense that depending on the direction of the magnetic field they propagate in one direction only at a given edge---independent of spin [see Fig.~\ref{fig:Scheme}~(c)].

Following the experimental demonstration of the QSH effect in HgTe-based QWs, much effort has been invested in the theoretical investigation of the properties of 2D topological insulators, their helical edge states, and possible applications.\cite{Zhou2008:PRL,Krueckl2011:PRL,Beugeling2012:PRB,Reinthaler2012:PRB,Lunde2013:PRB,Pikulin2014:PRB,Hofer2014:EPL,Geissler2014:PRB,Amaricci2015:PRL} At the heart of the QSH state are relativistic corrections, which can---if strong enough---result in band inversion,\cite{Chadi1972:PRB,Zhu2012:PRB} that is, a situation where the normal order of the conduction and valence bands is inverted and which can lead to peculiar effects such as the formation of interface states.\cite{Dyakonov1981:JETPL,Volkov1985:JETPL,Pankratov1987:SolidStateCommunications} In HgTe/CdTe QWs, the band structure of the 2D system formed in the well is inverted if the thickness $d$ of the HgTe QW exceeds the critical thickness $d_\mathrm{c}\approx6.3$ nm, whereas the band structure is normal for $d<d_\mathrm{c}$. Hence, at $B=0$, QSH states are absent in HgTe QWs with $d<d_\mathrm{c}$, but can appear for $d>d_\mathrm{c}$.

\begin{figure}[t]
\centering
\includegraphics*[width=8.65cm]{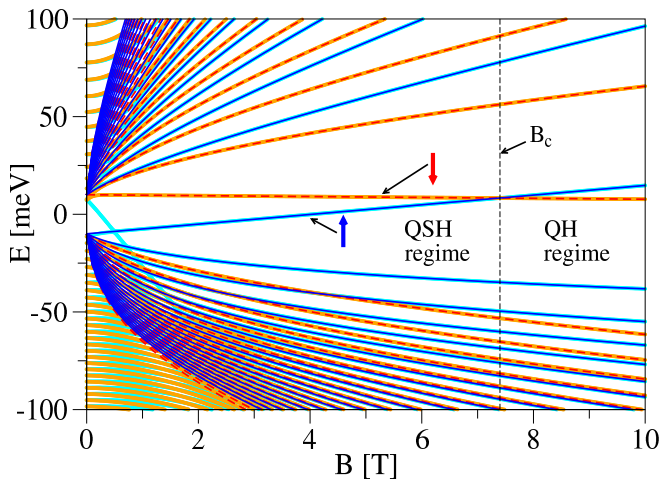}
\caption{(Color online) Magnetic field dependence of the states at $k=0$ in a finite strip of width $w=200$ nm compared to the bulk LLs. The thinner solid and dashed lines represent bulk LLs for $s=\uparrow$ and $s=\downarrow$, respectively. The levels of the finite-strip geometry are displayed by thick lines. All levels displayed here have been calculated for band parameters corresponding to $d=7.0$ nm, where the effective model introduced in Ref.~\onlinecite{Bernevig2006:Science} yields a critical magnetic field $B_\mathrm{c}\approx7.4$ T. From Ref.~\onlinecite{Scharf2012:PRB2}.}\label{fig:Landau_levels}
\end{figure}

If a magnetic field is applied to an inverted HgTe QW, Landau levels (LLs) and their accompanying edge states arise. It has been known for a long time that below a critical magnetic field $B_\mathrm{c}$ the uppermost valence LL has electron-like character and the lowest conduction LL has hole-like character in these inverted QW structures.\cite{Meyer1990:PRB,Truchsess1997:HMFPS,Schultz1998:PRB} In this situation, counterpropagating, spin-polarized states also still exist. Thus, the QSH state persists even at finite magnetic field $B<B_\mathrm{c}$, although these approximate QSH states are no longer protected by time-reversal symmetry.\cite{Tkachov2010:PRL,Tkachov2012:PhysicaE,Chen2012:PRB,Scharf2012:PRB2} We will call this inverted regime of $B<B_\mathrm{c}$, where both QSH and QH states coexist, the QSH regime in the following. For larger magnetic fields $B>B_\mathrm{c}$, the band ordering becomes normal (see Fig.~\ref{fig:Landau_levels}) and only QH states can be found. This normal regime, that is, a HgTe QW with $d>d_\mathrm{c}$ and $B>B_\mathrm{c}$ or a HgTe QW with $d<d_\mathrm{c}$, will be called the QH regime throughout the paper.

There has been a great amount of experimental interest in magneto-transport and (magneto)-optical properties of HgTe-based QWs\cite{Koenig2007:Science,Buettner2011:NatPhys,Gusev2013:PRL,Gusev2013:PRB,Kvon2014:JETP} and other topological insulators:\cite{Kim2012:NatPhys,Kim2013:NatComm,Aguilar2012:PRL,Costache2014:PRL,Lee2014:PRB,Chapler2014:PRB} For example, magneto-transport measurements near the charge neutrality point in a system which contains electrons and holes point to so-called snake states, known also from other materials,\cite{Badalyan2001:PRB,Shakouri2013:PRB} playing an important role in this regime.\cite{Gusev2010:PRL} Terahertz experiments have revealed a giant magneto-optical Faraday effect in HgTe thin films.\cite{Shuvaev2011:PRL} Furthermore, a resonant photocurrent has been observed in terahertz experiments on HgTe QWs with critical thickness and argued to originate from the cyclotron resonance of a linear Dirac dispersion.\cite{Olbrich2013:PRB,Zoth2014:PRB} Spectroscopic measurements also point to a LL spectrum characteristic of a Dirac system.\cite{Ludwig2014:PRB}

While there are theoretical studies on magneto-transport,\cite{Tkachov2011:PRB,Culcer2011:PRB} anomalous galvanomagnetism,\cite{Tkachov2011:PRB2} and magneto-optical effects such as the Faraday and Kerr effects\cite{Tse2010:PRL,Tse2010:PRB,Li2013:PRB,Li2015:PRB} in 2D topological insulators and topological insulator films, our goal here is to investigate magneto-optical properties of HgTe/CdTe QWs and search for signatures of the transition between the QSH and QH regimes in inverted QWs.

The paper is organized as follows: Following the introduction of the model and formalism in Sec.~\ref{Sec:Model}, the results for a bulk HgTe QW are discussed in Sec.~\ref{Sec:Bulk}, while Sec.~\ref{Sec:Finite_Strip} contains the main focus of this work and is devoted to the discussion of finite strip geometries. A brief summary concludes the paper.

\section{Model and methods}\label{Sec:Model}

\subsection{Model system}\label{Sec:ModelSystem}

For a description of the HgTe/CdTe QWs situated in the $xy$-plane and subject to a perpendicular magnetic field $\mathbf{B}=B\mathbf{e}_z$ (with $B>0$ throughout this paper), we use the gauge $\mathbf{A}(\mathbf{r})=-By\mathbf{e}_x$ for the magnetic vector potential, which is convenient if the system investigated is confined in the $y$-direction. Then, the QW is governed by the 2D, effective $4\times4$ Hamiltonian\cite{Bernevig2006:Science,Koenig2008:JPSJ}
\begin{equation}\label{SpinPolarized_Hamiltonian}
\begin{aligned}
\hat{H}_0=&\mathcal{C}\mathbf{1}_4+\mathcal{M}\Gamma_5-\frac{\mathcal{D}\mathbf{1}_4+\mathcal{B}\Gamma_5}{\hbar^2}\left[\left(\hat{p}_x-\frac{\hbar y}{l_{\mathsmaller{B}}^2}\right)^2+\hat{p}_y^2\right]\\
&+\frac{\mathcal{A}\Gamma_1}{\hbar}\left(\hat{p}_x-\frac{\hbar y}{l_{\mathsmaller{B}}^2}\right)+\frac{\mathcal{A}\Gamma_2}{\hbar}\hat{p}_y+\frac{\mu_{\mathsmaller{B}}B\Gamma^z_g}{2},
\end{aligned}
\end{equation}
where $\hat{p}_x$ and $\hat{p}_y$ are the momentum operators, $\mathcal{A}$, $\mathcal{B}$, $\mathcal{C}$, $\mathcal{D}$, and $\mathcal{M}$ are material parameters depending on the QW thickness $d$ (along the $z$-direction), $l_{\mathsmaller{B}}=\sqrt{\hbar/e|B|}=\sqrt{\hbar/eB}$ and $\mu_{\mathsmaller{B}}$ denote the magnetic length and the Bohr magneton, respectively, $e=|e|$ is the elementary charge, and $\hbar$ is Planck's constant. The effective Hamiltonian~(\ref{SpinPolarized_Hamiltonian}) captures the essential physics in HgTe/CdTe QWs at low energies and describes the spin-polarized electron-like ($E$) and heavy hole-like ($H$) states $\ket{E\uparrow}$, $\ket{H\uparrow}$, $\ket{E\downarrow}$, and $\ket{H\downarrow}$ near the $\Gamma$ point.

The matrices in Eq.~(\ref{SpinPolarized_Hamiltonian}) are given by the $4\times4$ unity matrix $\mathbf{1}_4$ and
\begin{equation}\label{matrices_SpinPolarized}
\begin{aligned}
\Gamma_1=\left(\begin{array}{cc}
         \sigma_x & 0 \\
         0 & -\sigma_x \\
         \end{array}\right),\:
\Gamma_2=\left(\begin{array}{cc}
         -\sigma_y & 0 \\
         0 & -\sigma_y \\
         \end{array}\right),\\
\Gamma_5=\left(\begin{array}{cc}
         \sigma_z & 0 \\
         0 & \sigma_z \\
         \end{array}\right),\:
\Gamma^z_g=\left(\begin{array}{cc}
         \sigma_g & 0 \\
         0 & -\sigma_g \\
         \end{array}\right),
\end{aligned}
\end{equation}
where $\sigma_x$, $\sigma_y$, and $\sigma_z$ denote the Pauli matrices describing electron- and hole-like states ($E/H$). Likewise, $\sigma_g=\diag(g_{\mathsmaller{\mathrm{e}}},g_{\mathsmaller{\mathrm{h}}})$ is a $2\times2$-matrix in the space spanned by $E$ and $H$ and contains the effective (out-of-plane) g-factors $g_{\mathsmaller{\mathrm{e}}}$ and $g_{\mathsmaller{\mathrm{h}}}$ of the $E$ and $H$ bands, respectively. Like the other material parameters, the g-factors depend on the thickness $d$ of the QW.\cite{Koenig2008:JPSJ,Buettner2011:NatPhys} Whether the QW is in the normal or inverted regime, is determined by $\mathcal{M}$ and $\mathcal{B}$: If $\mathcal{M}/\mathcal{B}<0$, the band structure is normal, whereas for a QW thickness $d>d_\mathrm{c}$, the band structure is inverted and $\mathcal{M}/\mathcal{B}>0$.

Moreover, we also investigate the effect of spin-orbit coupling (SOC) corrections, which---to lowest order---are described by\cite{Rothe2010:NJoP,Buettner2011:NatPhys}
\begin{equation}\label{SOC_Hamiltonian}
\hat{H}_\mathrm{SOC}=\Delta_0\Gamma_\mathrm{BIA}+\frac{\xi_e}{\hbar}\left[\Gamma_\mathrm{SIA1}\left(\hat{p}_x-\frac{\hbar y}{l_{\mathsmaller{B}}^2}\right)+\Gamma_\mathrm{SIA2}\hat{p}_y\right],
\end{equation}
where the first term describes the bulk inversion asymmetry of HgTe with its magnitude $\Delta_0$, the remaining terms are the leading-order contribution to the structural inversion asymmetry due to the QW potential\cite{Zutic2004:RMP,Fabian2007:APS} with the coefficient $\xi_e$, and the matrices are given by
\begin{equation}\label{matrices_SpinUnpolarized}
\begin{aligned}
\Gamma_\mathrm{SIA1}=\left(\begin{array}{cc}
         0 & \i\sigma_p \\
         -\i\sigma_p & 0 \\
         \end{array}\right),\:
\Gamma_\mathrm{SIA2}=\left(\begin{array}{cc}
         0 & \sigma_p \\
         \sigma_p & 0 \\
         \end{array}\right),\\
\Gamma_\mathrm{BIA}=\left(\begin{array}{cc}
         0 & -\i\sigma_y \\
         \i\sigma_y & 0 \\
         \end{array}\right),
\end{aligned}
\end{equation}
with $\sigma_p=\diag(1,0)$.

In this paper, we consider two geometries for the system described by Eqs.~(\ref{SpinPolarized_Hamiltonian})-(\ref{matrices_SpinUnpolarized}): (i) bulk, that is, an infinite system in the $xy$-plane, and---as the main focus of this work---(ii) a finite strip with the width $w$ in the $y$-direction. For both cases, we apply periodic boundary conditions in the $x$-direction, and the confinement in case (ii) can be described by adding the infinite hard-wall potential
\begin{equation}\label{confining_potential_finite_strip}
\begin{aligned}
V(y)=\left\{\begin{array}{ll}
            0 & \mathrm{for}\quad \left|y\right|<w/2\\
	    \infty & \mathrm{elsewhere.}
            \end{array}\right.
\end{aligned}
\end{equation}
Thus, the total Hamiltonian reads as
\begin{equation}\label{Total_Hamiltonian}
\hat{H}=\hat{H}_0+\hat{H}_\mathrm{SOC}+V(y)\mathbf{1}_4
\end{equation}
if SOC and confinement are taken into account.

In both cases, (i) and (ii), translational invariance along the $x$-direction is preserved by the Hamiltonian~(\ref{Total_Hamiltonian}). Thus, the wave vector $k$ in the $x$-direction is a good quantum number. Without SOC, that is, for $\Delta_0=0$ and $\xi_e=0$, spin is also a good quantum number and the spin-polarized eigenstates read as
\begin{equation}\label{SpinPolarized_states}
\Psi^s_{nk}(\mathbf{r})=\frac{\e^{\i kx}}{\sqrt{L}}\left(\begin{array}{c} f^s_{nk}(y)\\ g^s_{nk}(y)\\ \end{array}\right)\otimes\chi_s,
\end{equation}
where $n$ is a band index, $s$ the spin quantum number with its respective spinor $\chi_s$, $L$ the length of the strip in the $x$-direction, and the functions $f^s_{nk}(y)$ and $g^s_{nk}(y)$ as well as the corresponding energy $\epsilon_{ns}(k)$ can be determined numerically or analytically for both cases (i) and (ii) from the respective Schr\"{o}dinger equations (see the Appendix~\ref{Sec:AppendixEigenspectrum} and Ref.~\onlinecite{Scharf2012:PRB2} for explicit solutions). If SOC is taken into account, the eigenstates are still translationally invariant, but no longer spin-polarized and thus are given by
\begin{equation}\label{SpinUnpolarized_states}
\Psi_{nk}(\mathbf{r})=\frac{\e^{\i kx}}{\sqrt{L}}\left(\begin{array}{c} f^1_{nk}(y)\\g^1_{nk}(y)\\f^2_{nk}(y)\\g^2_{nk}(y) \end{array}\right),
\end{equation}
where $n$ is again a band index, and we determine the functions $f^1_{nk}(y)$, $g^1_{nk}(y)$, $f^2_{nk}(y)$, and $g^2_{nk}(y)$ as well as the corresponding energy $\epsilon_{n}(k)$ numerically with a finite-difference scheme. The eigenstates given by Eqs.~(\ref{SpinPolarized_states}) and~(\ref{SpinUnpolarized_states}) and their corresponding eigenenergies can then be used to calculate the magneto-optical conductivity via the Kubo formalism as will be discussed in the next section.

\subsection{Kubo formula for the magneto-optical conductivity}\label{Sec:ModelKubo}

Applying standard linear-response theory, one can write down Kubo formulas for the (magneto-)optical conductivities
\begin{equation}\label{OpticalConductivity}
\sigma_{lm}(\omega)=\frac{\i\Pi^R_{lm}(\omega)}{S\hbar\omega},
\end{equation}
where the retarded current-current correlation function $\Pi^R_{lm}(\omega)$ can be determined from the imaginary-time correlation function
\begin{equation}\label{CurrentCurrentCorrelationFunction}
\Pi_{lm}\left(\i\omega_n\right)=-\int\limits_{0}^{\hbar\beta}\d\tau\left\langle\mathcal{T}\left[\hat{I}_l(\tau)\hat{I}_m(0)\right]\right\rangle\e^{\i\omega_n\tau}
\end{equation}
via the formula $\Pi_{lm}^R\left(\omega\right)=\Pi_{lm}\left(\omega+\i0^\mathsmaller{+}\right)$ and $l$ and $m$ denote the $x$- or $y$-directions.\cite{Mahan2000,FetterWalecka2003,BruusFlensberg2006} Here, $\hat{I}_l(\tau)$ denotes the charge current operator derived from the Hamiltonian~(\ref{Total_Hamiltonian}) as described in the Appendix~\ref{Sec:AppendixCurrent}, $S$ the area of the QW, $\i\omega_n$ a bosonic frequency, $\tau$ an imaginary time, $\mathcal{T}$ the imaginary time-ordering operator, $\langle...\rangle$ the thermal average, and $\beta=1/(k_\mathrm{B}T)$ with the temperature $T$ and the Boltzmann constant $k_\mathrm{B}$.

Hence, we are left with the calculation of the retarded current-current correlation function, which can be determined from Eq.~(\ref{CurrentCurrentCorrelationFunction}). In this paper, we investigate a simple model: We assume that scattering by impurities can be described by a constant, phenomenological scattering rate $\Gamma/\hbar$ and do not explicitly consider any other processes such as, for example, electron-phonon coupling.\footnote{The exact shape of the absorption peaks shown in Sec.~\ref{Sec:Bulk} and~\ref{Sec:Finite_Strip} would depend on the details of the scattering mechanism. However, using a constant scattering rate $\Gamma/\hbar$ still gives a good, qualitative description of the essential mechanisms relevant for magneto-optical experiments.} Next, we introduce the spectral function, which in the spin-polarized case, that is, for $\Delta_0=0$ and $\xi_e=0$, is given by
\begin{equation}\label{SpectralFunction}
\mathcal{A}_{ns}(k,\omega)=\frac{2\hbar\Gamma}{\left[\hbar\omega-\epsilon_{ns}(k)+\mu\right]^2+\Gamma^2},
\end{equation}
where $\epsilon_{ns}(k)$ is the energy of the eigenstate labeled by the quantum numbers $n$, $k$, and $s$ of the system and $\mu$ is the chemical potential.

If we insert the current operator in Eq.~(\ref{CurrentCurrentCorrelationFunction}), express the Green's functions in the resulting equation with the help of the spectral function~(\ref{SpectralFunction}), calculate the sum over bosonic frequencies, and integrate over the resulting Dirac-$\delta$ functions, we obtain the magneto-optical conductivity tensors
\begin{equation}\label{sigmaxx}
\begin{aligned}
\mathrm{Re}\left[\sigma_{ll}(\omega)\right]=&\frac{\sigma_0}{4\pi S\omega}\sum\limits_{n,n',k,s}\left|d_{nn'}^{l,s}(0,k)\right|^2\\
&\times\int\d\omega'\mathcal{A}_{ns}(k,\omega')\mathcal{A}_{n's}(k,\omega+\omega')\\
&\times\left[n_\mathrm{FD}(\hbar\omega')-n_\mathrm{FD}(\hbar\omega'+\hbar\omega)\right]
\end{aligned}
\end{equation}
and
\begin{equation}\label{sigmaxy}
\begin{aligned}
\mathrm{Im}\left[\sigma_{xy}(\omega)\right]=&\frac{\sigma_0}{4\pi S\omega}\sum\limits_{n,n',k,s}\mathrm{Im}\left\{d_{nn'}^{x,s}(0,k)\left[d_{nn'}^{y,s}(0,k)\right]^*\right\}\\
&\times\int\d\omega'\mathcal{A}_{ns}(k,\omega')\mathcal{A}_{n's}(k,\omega+\omega')\\
&\times\left[n_\mathrm{FD}(\hbar\omega')-n_\mathrm{FD}(\hbar\omega'+\hbar\omega)\right]
\end{aligned}
\end{equation}
where $n_\mathrm{FD}(\epsilon)=1/\left[\exp(\beta\epsilon)+1\right]$ and $\sigma_0=e^2/\hbar$.\cite{Mahan2000,FetterWalecka2003,BruusFlensberg2006} Equations~(\ref{sigmaxx}) and~(\ref{sigmaxy}) also contain the dipole matrix elements $d_{nn'}^{l,s}(0,k)$ for transitions between bands $n$ and $n'$, where spin $s$ and momentum $k$ are conserved and which can be obtained from the current operator $\hat{I}_l$ and the spin-polarized eigenstates in Eq.~(\ref{SpinPolarized_states}) as detailed in the Appendix~\ref{Sec:AppendixCurrent}. Finally, the conductivities $\mathrm{Im}\left[\sigma_{ll}(\omega)\right]$ and $\mathrm{Re}\left[\sigma_{xy}(\omega)\right]$ are then calculated from Eqs.~(\ref{sigmaxx}) and~(\ref{sigmaxy}) by using Kramers-Kronig relations.

Equations~(\ref{SpectralFunction})-(\ref{sigmaxy}) are written explicitly for the case, where spin is a good quantum number and optical transitions are only permitted between states with the same spin quantum number. Spin conductivities $\sigma^\mathrm{diff}_{lm}(\omega)$ can then also be defined by replacing the sum $\sum_s$ by $\sum_ss$ in Eqs.~(\ref{sigmaxx}) and~(\ref{sigmaxy}). In case SOC is considered and spin is no longer a good quantum number, the conductivities can be calculated in a similar way. Then, the equations determining the conductivities are given by Eqs.~(\ref{SpectralFunction})-(\ref{sigmaxy}), but with the spin indices and the summation over spin omitted. Furthermore, the spin-unpolarized dipole matrix elements $d_{nn'}^{l}(0,k)$ for momentum-conserving transitions between bands $n$ and $n'$ obtained from the current operator $\hat{I}_l$ and the spin-unpolarized eigenstates~(\ref{SpinUnpolarized_states}) have to be used.

In the following, we will use Eqs.~(\ref{SpectralFunction})-(\ref{sigmaxy}) and their spin-unpolarized generalizations to calculate the magneto-optical conductivities for geometries (i), that is, bulk, and (ii), that is, a finite strip.

\section{Bulk}\label{Sec:Bulk}

As an introduction to the basic magneto-optical properties of HgTe QWs, we first investigate the magneto-optical conductivity in a bulk HgTe QW without SOC, that is, where $\Delta_0=0$ and $\xi_e=0$ and spin $s$ is a good quantum number. In this case, there is no confining potential $V(y)$ in Eq.~(\ref{Total_Hamiltonian}) and the eigenenergies are given by the dispersionless and highly degenerate LLs $\epsilon_{ns}(k)\equiv\epsilon_{ns}$, where $n\in\mathbb{Z}$ is an integer. The LLs for $n=0$ read as
\begin{equation}\label{0LLup}
\epsilon_{0\uparrow}=\mathcal{C}+\mathcal{M}-\frac{\mathcal{D}+\mathcal{B}}{l_{\mathsmaller{B}}^2}+\frac{g_{\mathsmaller{\mathrm{e}}}}{2}\mu_{\mathsmaller{B}}B
\end{equation}
and
\begin{equation}\label{0LLdown}
\epsilon_{0\downarrow}=\mathcal{C}-\mathcal{M}-\frac{\mathcal{D}-\mathcal{B}}{l_{\mathsmaller{B}}^2}-\frac{g_{\mathsmaller{\mathrm{h}}}}{2}\mu_{\mathsmaller{B}}B,
\end{equation}
while for $n\neq0$ they read as
\begin{equation}\label{LLup}
\begin{aligned}
\epsilon_{n\uparrow}&=\mathcal{C}-\frac{2\mathcal{D}|n|+\mathcal{B}}{l_{\mathsmaller{B}}^2}+\frac{g_{\mathsmaller{\mathrm{e}}}+g_{\mathsmaller{\mathrm{h}}}}{4}\mu_{\mathsmaller{B}}B+\mathrm{sgn}(n)\\
&\times\sqrt{\frac{2|n|\mathcal{A}^2}{l_{\mathsmaller{B}}^2}+\left(\mathcal{M}-\frac{2\mathcal{B}|n|+\mathcal{D}}{l_{\mathsmaller{B}}^2}+\frac{g_{\mathsmaller{\mathrm{e}}}-g_{\mathsmaller{\mathrm{h}}}}{4}\mu_{\mathsmaller{B}}B\right)^2}
\end{aligned}
\end{equation}
and
\begin{equation}\label{LLdown}
\begin{aligned}
\epsilon_{n\downarrow}&=\mathcal{C}-\frac{2\mathcal{D}|n|-\mathcal{B}}{l_{\mathsmaller{B}}^2}-\frac{g_{\mathsmaller{\mathrm{e}}}+g_{\mathsmaller{\mathrm{h}}}}{4}\mu_{\mathsmaller{B}}B+\mathrm{sgn}(n)\\
&\times\sqrt{\frac{2|n|\mathcal{A}^2}{l_{\mathsmaller{B}}^2}+\left(\mathcal{M}-\frac{2\mathcal{B}|n|-\mathcal{D}}{l_{\mathsmaller{B}}^2}-\frac{g_{\mathsmaller{\mathrm{e}}}-g_{\mathsmaller{\mathrm{h}}}}{4}\mu_{\mathsmaller{B}}B\right)^2}
\end{aligned}
\end{equation}
for spin-up and -down electrons, respectively.\cite{Buettner2011:NatPhys,Scharf2012:PRB2} Here, $n<0$ and $n>0$ denote valence and conduction LLs, respectively, while of the two LLs at $n=0$, one belongs to the conduction band and the other to the valence band.\footnote{Which spin quantum number at $n=0$ belongs to the conduction or valence bands depends on the QW thickness and the magnetic field. In the QSH regime, the spin-up zero LL lies below the spin-down zero LL.} The critical magnetic field, where those two zero levels cross and which separates the QSH and QH regimes, can be calculated from the condition $\epsilon_{0\uparrow}=\epsilon_{0\downarrow}$ and yields\cite{Scharf2012:PRB2}
\begin{equation}\label{B_crossing}
B_\mathrm{c}=\frac{\mathcal{M}}{e\mathcal{B}/\hbar-\left(g_{\mathsmaller{\mathrm{e}}}+g_{\mathsmaller{\mathrm{h}}}\right)\mu_\mathsmaller{B}/4}.
\end{equation}
Calculating the dipole matrix elements $d_{nn'}^{l,s}(0,k)$ from the corresponding eigenstates yields the optical selection rules
\begin{equation}\label{BulkSelectionRules}
|n|\to|n\pm1|,\,s\to s,\,k\to k
\end{equation}
for transitions between LLs. 

\begin{figure}[t]
\centering
\includegraphics*[width=8.65cm]{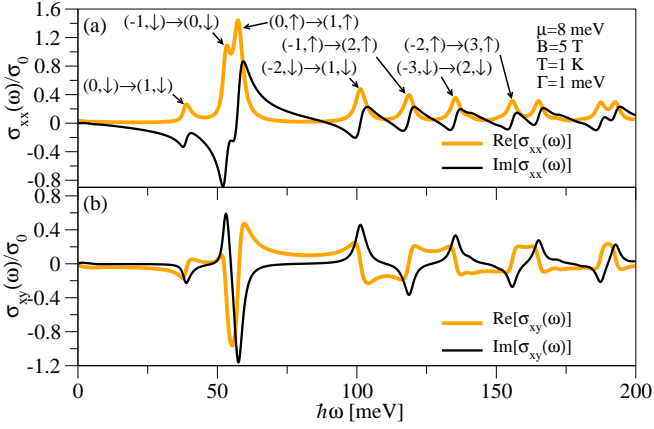}
\caption{(Color online) Real and imaginary parts of the (a) longitudinal magneto-optical and (b) optical Hall conductivities $\sigma_{xx}(\omega)$ and $\sigma_{xy}(\omega)$ in a bulk HgTe QW of thickness $d=7.0$ nm with some transitions explicitly labeled.}\label{fig:Bulk_general}
\end{figure}

To illustrate the resulting absorption spectrum, Fig.~\ref{fig:Bulk_general} shows the real and imaginary parts of the numerically\footnote{In all the numerical results for the conductivity presented in this paper, the integrals over the frequencies have been calculated on a discrete one-dimensional grid with $\Delta(\hbar\omega)=0.1$ meV. If $\mathrm{Im}\left[\sigma_{xx}(\omega)\right]$ or $\mathrm{Re}\left[\sigma_{xy}(\omega)\right]$ are determined from Kramers-Kronig relations, we use an upper cutoff for the integrals which is 300 meV above the maximal value of $\hbar\omega$ displayed. Thus, $\mathrm{Re}\left[\sigma_{xx}(\omega)\right]$ or $\mathrm{Im}\left[\sigma_{xy}(\omega)\right]$ need to be computed up to this upper cutoff. In this way, we ensure that the results for $\mathrm{Im}\left[\sigma_{xx}(\omega)\right]$ or $\mathrm{Re}\left[\sigma_{xy}(\omega)\right]$ are converged for the values of $\omega$ shown.} obtained magneto-optical conductivities $\sigma_{xx}(\omega)$ and $\sigma_{xy}(\omega)$ for the material parameters $\mathcal{A}=364.5$ meV nm, $\mathcal{B}=-686.0$ meV nm$^2$, $\mathcal{C}=0$, $\mathcal{D}=-512.0$ meV nm$^2$, $\mathcal{M}=-10.0$ meV, $g_\mathsmaller{\mathrm{e}}=22.7$, and $g_\mathsmaller{\mathrm{h}}=-1.21$, corresponding to a QW thickness $d=7.0$ nm $>d_\mathrm{c}$,\cite{Koenig2008:JPSJ,Qi2011:RMP} that is, for parameters in the QSH regime (at $B=0$). The magnetic field is chosen to be $B=5$ T, that is, a magnetic field below $B_\mathrm{c}\approx7.4$ T for which the model still shows an inverted band structure (see Fig.~\ref{fig:Landau_levels}), while the remaining parameters are chosen to be $T=1$ K, $\mu=8$ meV, and $\Gamma=1$ meV. Moreover, we note that the remaining components of the conductivity tensor can be determined from $\sigma_{yy}(\omega)=\sigma_{xx}(\omega)$ and $\sigma_{xy}(\omega)=-\sigma_{yx}(\omega)$ for the bulk system considered in this section.

As can be seen in Figs.~\ref{fig:Bulk_general}~(a) and~(b), there are multiple peaks in the absorptive components $\mathrm{Re}\left[\sigma_{xx}(\omega)\right]$ and $\mathrm{Im}\left[\sigma_{xy}(\omega)\right]$ corresponding to transitions between an occupied and an unoccupied LL state where the selection rules~(\ref{BulkSelectionRules}) have to be satisfied. Here, both intraband transitions, that is, transitions between only conduction LLs or only valence LLs, at low energies and interband transitions, that is, transitions between valence and conduction LLs, at higher energies are possible.

Due to the non-linear dependence of the LLs~(\ref{0LLup})-(\ref{LLdown}) on $n$, different allowed transitions between LLs have different energies resulting in the multiple absorption peaks shown in Fig.~\ref{fig:Bulk_general}~(a).\cite{Li2013:PRB} This behavior, also observed experimentally\cite{Ludwig2014:PRB} and reminiscent of the situation in graphene,\cite{Gusynin2007:PRL,Koshino2008:PRB,Pound2012:PRB,Scharf2013:PRB2} differs markedly from the behavior in a normal 2D electron gas, where equidistant LLs lead to only one absorption peak. Moreover, we find that, while every transition satisfying Eq.~(\ref{BulkSelectionRules}) occurs, different transitions are most probable for spin-up and spin-down LLs: If $n\geq1$, the squared dipole matrix element for a transition $-n\to n+1$ is typically five to ten times larger than the one of $-n\to n-1$ for spin-up LLs and vice versa for spin-down LLs. Consequently, the prominent interband transition peaks are governed by $-n\to n+1$ for spin-up LLs and by $-n\to n-1$ for spin-down LLs in Figs.~\ref{fig:Bulk_general}~(a) and~(b), while other interband transitions contribute much less to the absorption spectrum.

In addition to the absorptive components, the refractive components $\mathrm{Im}\left[\sigma_{xx}(\omega)\right]$ and $\mathrm{Re}\left[\sigma_{xy}(\omega)\right]$ are also displayed for completeness. Both, the absorptive and refractive components, are generally needed to calculate optical observables, such as reflection coefficients and angles, for example.

\begin{figure}[t]
\centering
\includegraphics*[width=8.65cm]{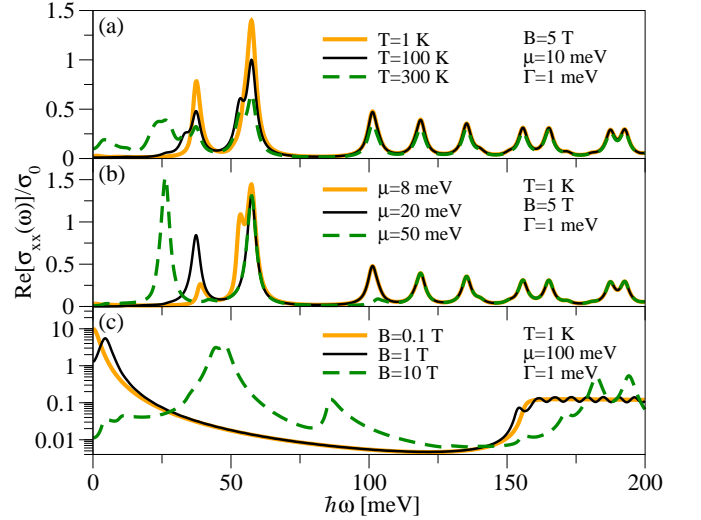}
\caption{(Color online) Real part of the longitudinal magneto-optical conductivity $\sigma_{xx}(\omega)$ in a bulk HgTe QW of thickness $d=7.0$ nm for different (a) temperatures $T$, (b) chemical potentials $\mu$, and (c) magnetic fields $B$.}\label{fig:Bulk_parameter_dependence}
\end{figure}

In Fig.~\ref{fig:Bulk_parameter_dependence}, the dependence of $\mathrm{Re}\left[\sigma_{xx}(\omega)\right]$ as a function of the frequency on different parameters is displayed for a bulk HgTe QW again in the QSH regime (at $B=0$). The temperature dependence is illustrated in Fig.~\ref{fig:Bulk_parameter_dependence}~(a), which shows $\mathrm{Re}\left[\sigma_{xx}(\omega)\right]$ for a fixed magnetic field, chemical potential, and broadening. Here, the main feature observed is that with increasing $T$, additional (intraband) transitions (forbidden at $T=0$) become more probable which then give rise to new peaks---mainly at low frequencies. For higher frequencies, on the other hand, the magneto-optical conductivity remains largely unaffected, although the interband peaks are slightly reduced as spectral weight is transferred to lower energies, while the total spectral weight is conserved.

Figure~\ref{fig:Bulk_parameter_dependence}~(b) shows $\mathrm{Re}\left[\sigma_{xx}(\omega)\right]$ for several different chemical potentials and a fixed magnetic field, temperature, and broadening. As $\mu$ increases, the intraband transition peaks move to lower energies, while the gap between intraband and interband transitions increases. Likewise, with decreasing $B$, the energy of the intraband transition peaks decreases and tends to zero, as can be seen in Fig.~\ref{fig:Bulk_parameter_dependence}~(c), which displays $\mathrm{Re}\left[\sigma_{xx}(\omega)\right]$ for several different values of $B$ and fixed $\mu$, $T$, and $\Gamma$.

The behavior of the intraband transition peaks with decreasing magnetic field or with increasing chemical potential can be qualitatively explained as originating from the LL spectrum in the vicinity of the Fermi level: For a fixed magnetic field, the LL spacing near the Fermi level decreases if the absolute value of the chemical potential is increased and thus situated in a denser region of the LL spectrum. Consequently, the energies of the intraband transitions are also decreased. On the other hand, with decreasing magnetic field the LL spacing decreases giving rise to lower energies of the intraband transitions. Moreover, the amplitudes of the interband peaks decrease with decreasing magnetic fields, as seen in Fig.~\ref{fig:Bulk_parameter_dependence}~(c). This can be interpreted as arising from optical transitions between increasingly denser regions of the LL spectrum, where the energy difference between different transitions is small compared to the broadening due to scattering.

%\begin{figure}[t]
%\centering
%\includegraphics*[width=8.65cm]{Fig5old}
%\caption{(Color online) Real part of the longitudinal magneto-optical conductivity $\sigma_{xx}(\omega_0)$ in a bulk HgTe QW of thickness $d=7.0$ nm as a function of $1/B$ for different frequencies $\omega_0$.}\label{fig:Bulk_iBfield_dependence}
%\end{figure}
%
%The dependence of the absorption on the magnetic field is also illustrated in Fig.~\ref{fig:Bulk_iBfield_dependence}, which shows $\mathrm{Re}\left[\sigma_{xx}(\omega_0)\right]$ as a function of the inverse magnetic field $1/B$ at a fixed temperature, chemical potential, broadening, and several different photon frequencies $\omega_0$.  Similar to its frequency dependence, with varying magnetic field, $\mathrm{Re}\left[\sigma_{xx}(\omega)\right]$ exhibits pronounced peaks for higher photon frequencies $\omega_0$.

\begin{figure}[t]
\centering
\includegraphics*[width=8.65cm]{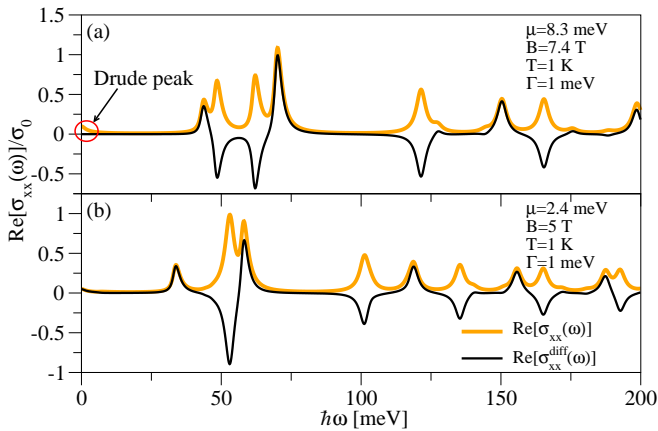}
\caption{(Color online) Real parts of the longitudinal and spin longitudinal magneto-optical conductivities, $\sigma_{xx}(\omega)$ and $\sigma^\mathrm{diff}_{xx}(\omega)$, respectively, in a bulk HgTe QW of thickness $d=7.0$ nm (a) at the critical magnetic field $B=B_\mathrm{c}\approx7.4$ T and (b) at $B=5$ T. In panel~(a), $\mu$ is chosen to be at the energy of the degenerate spin-up and spin-down zero mode Landau levels, while in panel~(b) $\mu$ is chosen to be at the energy of the spin-up zero mode Landau levels.}\label{fig:Bulk_comparison_spin}
\end{figure}

In the above discussion, we have investigated a QW with parameters corresponding to the topological regime (at $B=0$). However, the above conclusions on the behavior of the magneto-optical conductivity also apply to QWs with a thickness $d<d_\mathrm{c}$, that is, QWs in the topologically trivial regime. We conclude our discussion of bulk HgTe QWs by investigating also the spin magneto-optical conductivity $\sigma^\mathrm{diff}_{xx}(\omega)$, where one can indeed find a signature of the transition between the QSH and the QH regimes: As shown in Fig.~\ref{fig:Landau_levels}, the transition between the QSH regime at finite magnetic field and the QH regime in QWs with a thickness $d>d_\mathrm{c}$ occurs when the spin-up and spin-down zero LLs cross at a critical magnetic field $B_\mathrm{c}$.

If the sample is undoped as chosen in Fig.~\ref{fig:Bulk_comparison_spin}~(a), the chemical potential lies between the two zero LLs. This, however, means that for $B=B_\mathrm{c}$ the chemical potential will lie in the two degenerate zero LLs. Thus, there is an additional Drude transition peak at zero frequency as shown in Fig.~\ref{fig:Bulk_comparison_spin}~(a). This peak originates from two different transitions, namely spin-conserving transitions within the spin-up or spin-down zero LLs. As the squared dipole matrix elements for transitions within the spin-up and spin-down zero LLs are the same, both transitions contribute equally to the absorption peak at $\hbar\omega=0$. However, this means that there is no peak at $\hbar\omega=0$ for the spin conductivity $\sigma^\mathrm{diff}_{xx}(\omega)$ as spin-up and spin-down transitions cancel each other exactly.

This only happens at the critical field $B_\mathrm{c}$ in an undoped sample and is different from the situation if $\mu$ in a doped sample crosses one LL, an example of which is shown in Fig.~\ref{fig:Bulk_comparison_spin}~(b) for comparison. In this case, there is also a Drude peak at $\hbar\omega=0$ for $\sigma_{xx}(\omega)$, but since this peak arises from transitions within only one LL with a given spin quantum number, $\sigma^\mathrm{diff}_{xx}(\omega)$ also exhibits a peak at zero frequency and can thus be distinguished from the peak arising from the two zero LLs at $B_\mathrm{c}$.

\section{Finite Strip}\label{Sec:Finite_Strip}

While in the previous section we have investigated the magneto-optical properties of bulk HgTe QWs without SOC, which did not account for the presence of edge states and where we did not find significant differences between the QSH and the QH regimes, we will now turn to a finite-strip geometry described by the confining potential in Eq.~(\ref{confining_potential_finite_strip}). Since the finite-strip geometry contains boundaries, both chiral QH edge states as well as helical QSH edge states emerge at these boundaries under the appropriate conditions: If the width $w$ of the finite wire is large compared to $l_{\mathsmaller{B}}$, well localized QH edge states form in addition to the bulk LLs. In the QSH regime, that is, in a QW with $d>d_\mathrm{c}$ and below $B_\mathrm{c}$, there are also QSH edge states in addition to the QH states.

\begin{figure}[t]
\centering
\includegraphics*[width=8.65cm]{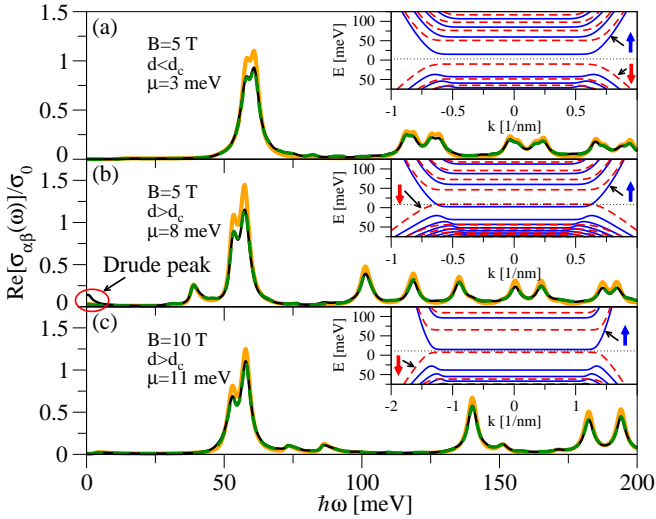}
\caption{(Color online) Real parts of the longitudinal magneto-optical conductivities $\sigma_{xx}(\omega)$ and $\sigma_{yy}(\omega)$ in (a) a HgTe QW of thickness $d=5.5$ nm and with a magnetic field $B=5$ T, (b) a HgTe QW with $d=7.0$ nm and $B=5$ T, and (c) a HgTe QW with $d=7.0$ nm and $B=10$ T. Here, spin-orbit coupling corrections are not included. The solid orange lines represent $\sigma_{xx}(\omega)=\sigma_{yy}(\omega)$ in a bulk system, while the solid black and the dashed green lines represent $\sigma_{xx}(\omega)$ and $\sigma_{yy}(\omega)$ in a finite strip of width $w=200$ nm. The insets in panels~(a)-(c) illustrate the respective energy spectra of the finite strip and the positions of the chemical potential (dotted lines).}\label{fig:FS_comparison}
\end{figure}

This can also be seen in Fig.~\ref{fig:FS_comparison}, which shows the real parts of the magneto-optical conductivities $\sigma_{xx}(\omega)$ and $\sigma_{yy}(\omega)$ for a finite-strip of width $w=200$ nm for different QW thicknesses and magnetic fields corresponding to different regimes\footnote{For $d>d_\mathrm{c}$, we use $d=7.0$ nm with the parameters given in Sec.~\ref{Sec:Bulk}, whereas we use parameters corresponding to $d=5.5$ nm for $d<d_\mathrm{c}$. In this case, the parameters are given as $\mathcal{A}=387$ meV nm, $\mathcal{B}=-480.0$ meV nm$^2$, $\mathcal{C}=0$, $\mathcal{D}=-306.0$ meV nm$^2$, and $\mathcal{M}=9.0$ in Ref.~\onlinecite{Qi2011:RMP}.} without SOC ($\Delta_0=0$ and $\xi_e=0$): a QW with $d<d_\mathrm{c}$, that is, a system with no QSH states at any magnetic field [Fig.~\ref{fig:FS_comparison}~(a)], a QW with $d>d_\mathrm{c}$ and $B<B_\mathrm{c}$, that is, a system possessing QSH states at the edges of the strip [Fig.~\ref{fig:FS_comparison}~(b)], and a QW with $d>d_\mathrm{c}$, but $B>B_\mathrm{c}$, that is, a system with no QSH states [Fig.~\ref{fig:FS_comparison}~(c)]. For each of the regimes, we have chosen $\mu$ to lie at the neutrality point between the uppermost valence LL and the lowest conduction LL. The insets of Figs.~\ref{fig:FS_comparison}~(a)-(c) show the respective energy spectra and the positions of $\mu$.

In the regimes without any QSH edge states [Figs.~\ref{fig:FS_comparison}~(a) and (c)], the band structure is gapped as every electron-like LL is a conduction band and situated above the hole-like valence bands. The edge states associated with an electron-like LL have positive curvature, while those associated with a hole-like LL have negative curvature.\cite{Badalyan2009:PRL} Figure~\ref{fig:FS_comparison}~(b), on the other hand, shows a different situation as the lowest electron-like (spin-up) LL is a valence band and the highest hole-like (spin-down) LL is a conduction band. Thus, there is a crossover between the dispersions of electron- and hole-like bands and one consequently finds counterpropagating, spin-polarized QSH states, in addition to QH edge states propagating in the same direction at a given boundary regardless of spin. For comparison, the magneto-optical conductivities for a bulk HgTe QW as investigated in Sec.~\ref{Sec:Bulk} are also displayed in Fig.~\ref{fig:FS_comparison} for the same magnetic fields and QW thicknesses $d$ as the finite-strip structures.

Compared to the situation in a bulk HgTe QW, we can see that in a finite strip there are still pronounced absorption peaks arising from the bulk LLs. However, the spectral weight of these peaks is reduced as transitions between edge states are also possible. Since the energy of these transitions can differ from the energy difference between two bulk LLs and the total spectral weight is conserved, some spectral weight is transferred from the peaks to the regions between the absorption peaks, a phenomenon clearly seen in Fig.~\ref{fig:FS_comparison}.

A second difference compared to the bulk system is that, even if the chemical potential is not situated in a bulk LL, there is a Drude-like absorption peak at low energies for $\sigma_{xx}(\omega)$. Analogous to the situation in Fig.~\ref{fig:Bulk_comparison_spin}, where a Drude peak at low energies originated from transitions within bulk LL states though, this absorption peak arises from transitions that occur when the energy of a QH or QSH edge state is at the chemical potential or close to it (within the broadening). This mechanism is present in both the QH and QSH regimes if the chemical potential is above or below the bulk band gap. The difference between those two regimes, however, is that edge states exist at every energy in the QSH regime, while there are no states with energies inside the bulk gap in the QH regime (see the insets in Fig.~\ref{fig:FS_comparison}). Hence, if in the QH regime the chemical potential lies inside the gap and the gap exceeds the energy scale associated with broadening, no low-energy absorption peak occurs as seen in Figs.~\ref{fig:FS_comparison}~(a) and~(c). Moreover, we note that this Drude peak appears only along the unconfined direction, but not for $\sigma_{yy}(\omega)$ along the confined direction. This is to be expected because only the $x$-direction is associated with free acceleration.

\begin{figure}[t]
\centering
\includegraphics*[width=8.65cm]{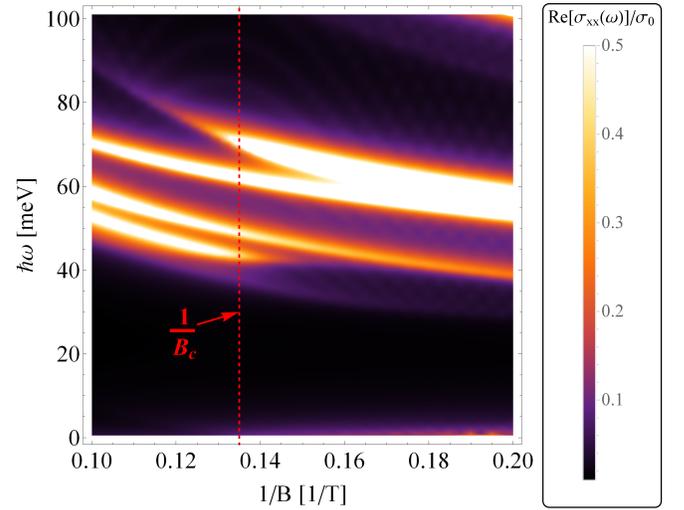}
\caption{(Color online) Real part of the total longitudinal magneto-optical conductivity $\sigma_{xx}(\omega)$ of a finite strip with width $w=200$ nm, $\Gamma=1$ meV, and chemical potential $\mu=8$ meV at temperature $T=1$ K as a function of the inverse magnetic field $1/B$ and $\hbar\omega$ for band parameters corresponding to a HgTe QW of thickness $d=7.0$ nm without spin-orbit coupling corrections. The vertical line indicates the inverse critical magnetic field separating the QSH and QH regimes and as calculated by Eq.~(\ref{B_crossing}).}\label{fig:FS_LowFrequencies_OmegaIB}
\end{figure}

The disappearance of the Drude peak if $\mu$ remains inside the bulk gap during the transition from the QSH to the QH regime is also illustrated in Fig.~\ref{fig:FS_LowFrequencies_OmegaIB}: Here, the absorption spectrum of a finite-strip geometry for a QW with $d=7.0$ nm $>d_\mathrm{c}$ (and without SOC) is shown as a function of $1/B$ and $\hbar\omega$ at low temperatures and fixed chemical potential. For $1/B=0.2$/T, the situation is described by Fig.~\ref{fig:FS_comparison}~(b). As $1/B$ is decreased, that is, as the magnetic field is increased, the absolute value of the LL energies and thus the photon energies of their associated absorption peaks increase. At low photon energies, there is a finite absorption, which vanishes for magnetic fields above the critical field $B_\mathrm{c}$ when the transition from the QSH into the QH regime occurs.

\begin{figure}[t]
\centering
\includegraphics*[width=8.65cm]{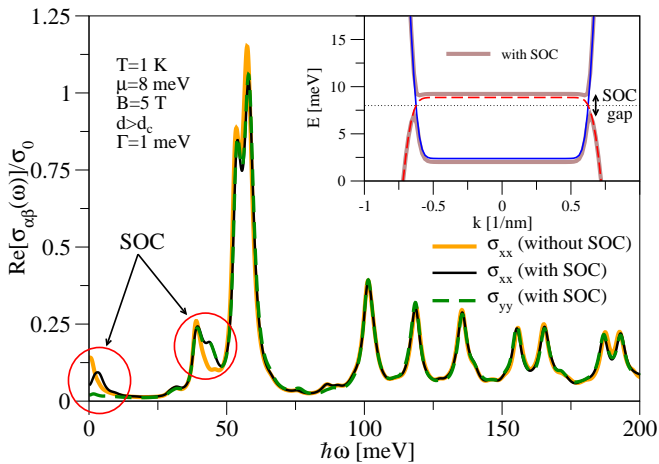}
\caption{(Color online) Real parts of the longitudinal magneto-optical conductivities $\sigma_{xx}(\omega)$ and $\sigma_{yy}(\omega)$ of a finite strip with width $w=200$ nm, a magnetic field $B=5$ T, and band parameters corresponding to a HgTe QW of thickness $d=7.0$ nm with and without spin-orbit coupling corrections as given in Eq.~(\ref{SOC_Hamiltonian}). The respective energy spectra and the position of the chemical potential (dotted line) are shown in the inset.}\label{fig:FS_Rashba}
\end{figure}

We now address the effect of SOC. While the states in Fig.~\ref{fig:FS_comparison} have been perfectly spin-polarized and characterized by the spin quantum number $s$, the situation changes if SOC is considered, that is, if $\Delta_0$ and/or $\xi_e$ are finite in the total Hamiltonian~(\ref{SOC_Hamiltonian}). This is illustrated in Fig.~\ref{fig:FS_Rashba}, where the real parts of the magneto-optical conductivities $\sigma_{xx}(\omega)$ and $\sigma_{yy}(\omega)$ are displayed for a QW of thickness $d=7.0$ nm $>d_\mathrm{c}$ at $B=5$ T (see above) with the additional SOC parameters $\Delta_0=1.6$ meV and $\xi_e=16.0$ meV nm.\cite{Buettner2011:NatPhys} The inset of Fig.~\ref{fig:FS_Rashba} compares the energy spectrum near the band gap in the presence of SOC with the spin-polarized energy spectrum in the absence of SOC. The most striking effect appears at the crossing between the spin-up and spin-down states, where a SOC gap is opened up. Consequently, the low-energy absorption peak present in the QSH regime without SOC is reduced if the chemical potential is situated inside this small gap. However, as long as the gap opened by SOC does not significantly exceed the broadening $\Gamma$, an increased absorption for $\sigma_{xx}(\omega)$ can still be observed in the QSH regime at low photon energies $\hbar\omega$. As bands farther away from the zero LLs are affected even less by SOC for the parameters given above, the absorption spectrum at higher energies remains nearly unaltered compared to spin-polarized case.

\begin{figure}[t]
\centering
\includegraphics*[width=8.65cm]{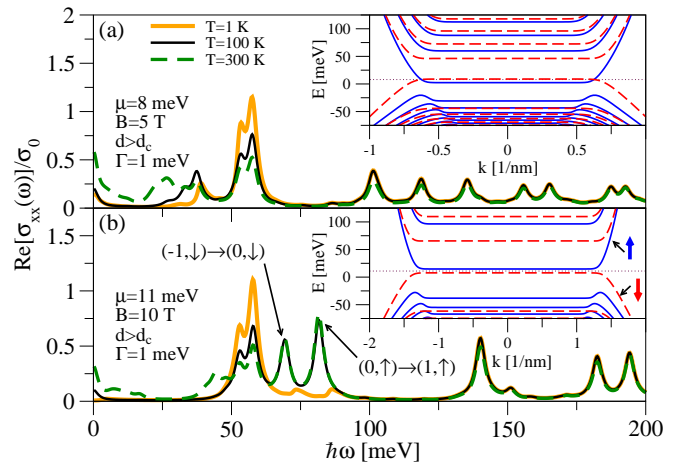}
\caption{(Color online) Real part of the longitudinal magneto-optical conductivity $\sigma_{xx}(\omega)$ of a finite strip with width $w=200$ nm, band parameters corresponding to a HgTe QW of thickness $d=7.0$ nm without spin-orbit coupling corrections, and magnetic fields of (a) $B=5$ T and (b) $B=10$ T for different temperatures $T$. The energy spectra and the positions of the chemical potential (dotted lines) are shown in the insets.}\label{fig:FS_temperature_dependence}
\end{figure}

\begin{figure}[t]
\centering
\includegraphics*[width=8.65cm]{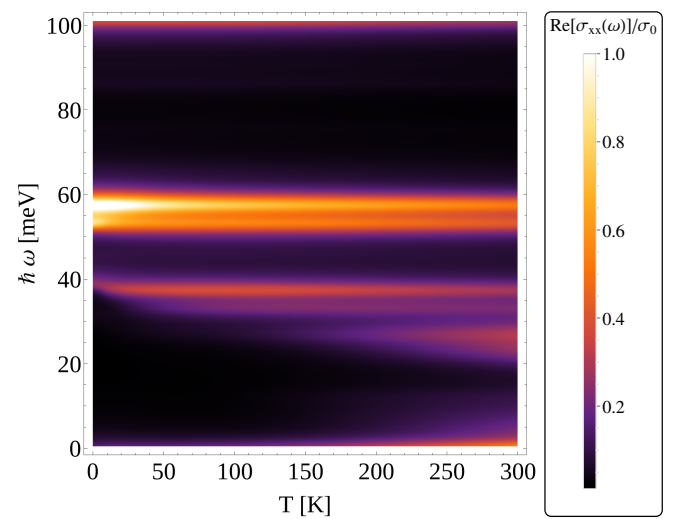}
\caption{(Color online) Real part of the total longitudinal magneto-optical conductivity $\sigma_{xx}(\omega)$ of a finite strip with width $w=200$ nm as a function of the temperature $T$ and $\hbar\omega$ for $B=5$ T, $\mu=8$ meV, $\Gamma=1$ meV, and band parameters corresponding to a HgTe QW of thickness $d=7.0$ nm without spin-orbit coupling corrections.}\label{fig:FS_OmegaT}
\end{figure}

Next, we turn our attention to the effect of temperature on the absorption spectrum in general and on the Drude-like peak arising from QSH states in particular and compare this to the QH regime. For this purpose, Figs.~\ref{fig:FS_temperature_dependence}~(a) and~(b) show $\mathrm{Re}\left[\sigma_{xx}(\omega)\right]$ in a finite-strip geometry (without SOC) at different temperatures for a QW with $d=7.0$ nm $>d_\mathrm{c}$ in the QSH and QH regimes, respectively, with the chemical potential situated between the zero LLs. Complementary, Fig.~\ref{fig:FS_OmegaT} displays $\mathrm{Re}\left[\sigma_{xx}(\omega)\right]$ for the setup of Fig.~\ref{fig:FS_temperature_dependence}~(a) as a function of $T$ and $\hbar\omega$.

Similar to the situation in a bulk QW illustrated in Fig.~\ref{fig:Bulk_parameter_dependence}~(a), additional transitions especially, but not exclusively, at low energies become more probable. This results in additional absorption peaks and an increase of the absorption at low energies, while spectral weight is transferred away from the intraband peaks as can be seen in Figs.~\ref{fig:FS_temperature_dependence}~and~\ref{fig:FS_OmegaT}. Since the spin-up and spin-down zero modes are very close to $\mu$ in these setups, additional transitions involving one of these LLs become especially likely if $T$ is increased and consequently the LL below $\mu$ is depopulated, while the LL above $\mu$ is populated. This is particularly striking in Fig.~\ref{fig:FS_temperature_dependence}~(b), where already at $T=100$ K pronounced peaks appear due to transitions $n=0\to1$ and $-1\to0$ for spin-up and spin-down LLs, respectively.

As the absorption at low energies becomes more probable, an enhancement of the low-energy absorption peak in the QSH regime can be observed in Figs.~\ref{fig:FS_temperature_dependence}~(a) and~\ref{fig:FS_OmegaT}. On the other hand, Fig.~\ref{fig:FS_temperature_dependence}~(b) shows that such a peak also appears in the QH regime with increasing temperature. The smaller the gap between conduction and valence LLs, the smaller is the temperature for the onset of this effect. Thus, at high $T$ both regimes exhibit a low-energy absorption peak for $\sigma_{xx}(\omega)$.

\begin{figure}[t]
\centering
\includegraphics*[width=8.65cm]{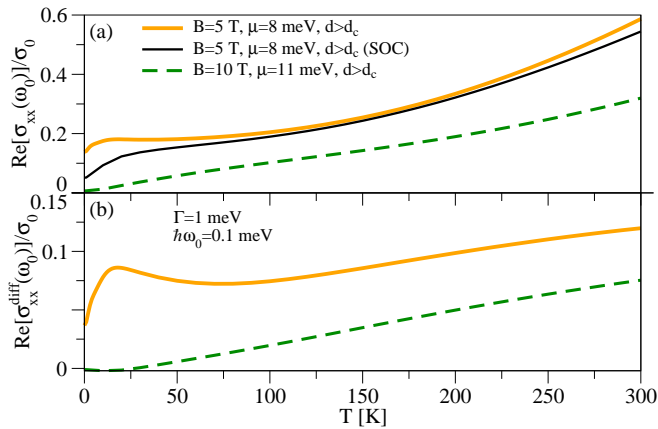}
\caption{(Color online) Real parts of the (a) total and (b) spin longitudinal magneto-optical conductivities $\sigma_{xx}(\omega)$ and $\sigma^\mathrm{diff}_{xx}(\omega)$ of a finite strip with width $w=200$ nm at low frequencies $\omega_0$ as a function of the temperature $T$ for different setups.}\label{fig:FS_LowFrequencies_T}
\end{figure}

This is also corroborated by Fig.~\ref{fig:FS_LowFrequencies_T}~(a), where the real part of $\sigma_{xx}(\omega)$ at low energies is displayed as a function of $T$ for a QW in the QSH regime, in the QSH regime with SOC, and in the QH regime with the chemical potential situated between the conduction and valence LLs. At low temperatures, there is a finite absorption in the QSH regime, even if reduced by SOC, whereas there is no absorption in the QH regime. With increasing $T$, the absorption also increases, in both the QSH and QH regimes, although the increase is more pronounced in the QSH regime. Moreover, SOC corrections are less pronounced at higher temperatures. Without SOC, a spin conductivity $\sigma^\mathrm{diff}_{xx}(\omega)$ can be defined, whose temperature dependence is shown in Fig.~\ref{fig:FS_LowFrequencies_T}~(b) and closely follows the temperature dependence of $\sigma_{xx}(\omega)$.

\begin{figure}[t]
\centering
\includegraphics*[width=8.65cm]{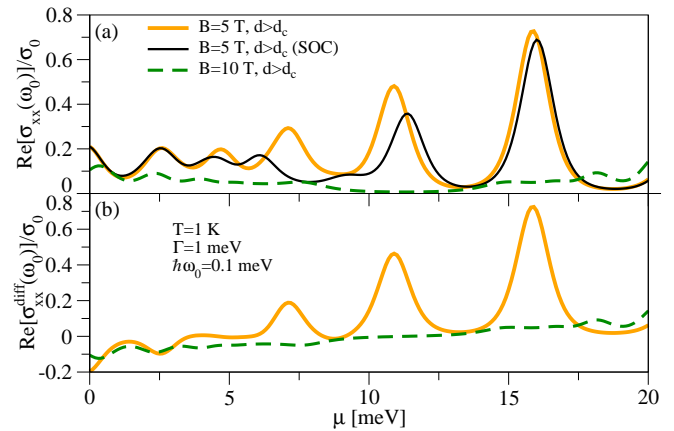}
\caption{(Color online) Real parts of the (a) total and (b) spin longitudinal magneto-optical conductivities $\sigma_{xx}(\omega)$ and $\sigma^\mathrm{diff}_{xx}(\omega)$, respectively, of a finite strip with width $w=200$ nm at low frequencies $\omega_0$ as a function of the chemical potential $\mu$ for different setups.}\label{fig:FS_LowFrequencies_mu}
\end{figure}

Finally, we study the dependence of the low-energy absorption on the chemical potential. For this purpose, Fig.~\ref{fig:FS_LowFrequencies_mu}~(a) shows the real parts of the conductivity $\sigma_{xx}(\omega)$ at low $\hbar\omega$ and $T$ as functions of $\mu$ for a QW in the QSH regime, in the QSH regime with SOC, and in the QH regime. The corresponding spin conductivities $\sigma^\mathrm{diff}_{xx}(\omega)$ for the regimes without SOC are displayed in Fig.~\ref{fig:FS_LowFrequencies_mu}~(b). As can be seen in Fig.~\ref{fig:FS_LowFrequencies_mu}~(a), the absorption in the QSH regime is finite and usually higher than in the QH regime. Spin-orbit coupling in the QSH regime only has a significant effect close to the crossing between the electron- and hole-like edge dispersions, that is, in the energy interval approximately between $5$ meV and $15$ meV. In this region, $\sigma_{xx}(\omega)$ is reduced due to SOC, although it does not vanish completely. The behavior with varying chemical potential in the QH regime, on the other hand, is different: If $\mu$ is situated in the gap between the valence and conduction LLs, there is no Drude-like peak and $\sigma_{xx}(\omega)$ vanishes.

\textit{Thus, by sweeping the gate-voltage and thereby varying the position of the chemical potential, it is possible to distinguish between the QSH and QH regimes on the basis of their their low-energy absorption as depicted in Fig.~\ref{fig:FS_LowFrequencies_mu}~(a).} We note that this behavior of the Drude peak in the absorption spectrum is consistent with and reflects the expected behavior of the dc conductance in the QSH and QH regimes.\cite{Bernevig2006:Science}

\begin{figure}[t]
\centering
\includegraphics*[width=8.65cm]{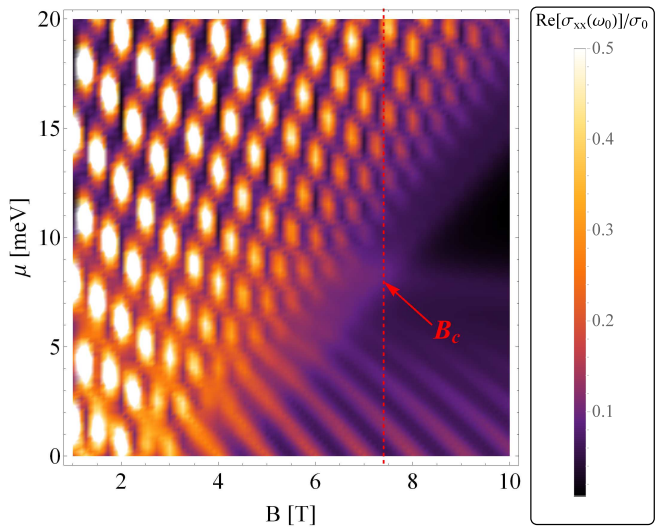}
\caption{(Color online) Real part of the total longitudinal magneto-optical conductivity $\sigma_{xx}(\omega)$ of a finite strip with width $w=200$ nm and $\Gamma=1$ meV at low energies $\hbar\omega_0=0.1$ meV and temperature $T=1$ K as a function of the chemical potential $\mu$ and magnetic field $B$ for band parameters corresponding to a HgTe QW of thickness $d=7.0$ nm without spin-orbit coupling corrections. The vertical line indicates the critical magnetic field separating the QSH and QH regimes and as calculated by Eq.~(\ref{B_crossing}).}\label{fig:FS_LowFrequencies_muB}
\end{figure}

This is also substantiated in Fig.~\ref{fig:FS_LowFrequencies_muB} which shows the low-energy absorption of a QW of thickness $d=7.0$ nm without SOC corrections as a function of both the chemical potential as well as the magnetic field. While in the QSH regime, $B<B_\mathrm{c}\approx7.4$ T, $\mathrm{Re}\left[\sigma_{xx}(\omega)\right]$ never vanishes entirely, there is a region where $\mathrm{Re}\left[\sigma_{xx}(\omega)\right]$ is completely suppressed in the QH regime above $B_\mathrm{c}$. With increasing magnetic field, the extent of this region grows as the gap between conduction and valence-LLs increases. Outside the region of suppressed conductance and low-energy absorption, the spectrum displays an oscillatory behavior as seen in Figs.~\ref{fig:FS_LowFrequencies_mu} and~\ref{fig:FS_LowFrequencies_muB}.

Finally, we remark that our description does not take into account many-body effects, such as (edge) magneto-plasmon resonances which can potentially play a prominent role for the absorption especially in narrow strips.\cite{Volkov1988:JETP,HansenSAS:1992,Mikhailov2004:PRB} Since the scheme that we propose to distinguish between the QSH and QH regimes is based on the dc conductivity, however, we think this scheme to be robust, even in the presence of edge magneto-plasmons, although the absorption at low, but finite energies can obtain a richer structure than the single-particle absorption calculated in this paper. Moreover, our single-particle calculations provide a useful benchmark against which to test experimental results.

\section{Conclusions}\label{Sec:Conclusions}

Inverted HgTe/CdTe QWs exhibit helical QSH states below a critical magnetic field $B_\mathrm{c}$, above which the band order is normal and only QH states can be found. In this work, we have studied the magneto-optical properties of such HgTe/CdTe QWs by calculating the magneto-optical conductivity using linear-response theory. We have considered both an infinite 2D system as well as a finite strip, that is, a 2D system that is confined in one direction. The normal and the inverted regimes both exhibit a series of pronounced absorption peaks corresponding to different transitions, both intraband as well as interband transitions, between non-equidistant LLs in the infinite system. Thus, it is hard to distinguish between the QSH and QH regimes based on these LL peaks.

We find that these bulk LL peaks also occur in the finite strip, where the presence of edge states results in an additional Drude-like absorption peak in $\sigma_{xx}(\omega)$ originating from low-energy transitions at the Fermi level. This Drude peak is always present in the QSH regime, while it vanishes in the QH regime if the chemical potential is situated in the bulk gap. If SOC corrections are included, we find that their effect is to open up a small gap in the QSH spectrum, which leads to a reduction, but not to a complete disappearance of this Drude peak. By sweeping the gate-voltage of a finite QW structure and thereby varying the position of the chemical potential, it is therefore possible to experimentally distinguish between the QSH and QH regimes on the basis of the stability (QSH) or disappearance (QH) of the Drude peak.

\acknowledgments
This work was supported by DFG Grants No. SCHA 1899/1-1 (B.S.) and SFB 689 (J.F.), U.S. ONR Grant No. N000141310754 (B.S., A.M.-A.), U.S. DOE, Office of Science BES, under Award DE-SC0004890 (I.\v{Z}), as well as the International Doctorate Program Topological Insulators of the Elite Network of Bavaria (J.F.).

\appendix

\section{Eigenspectrum and eigenstates}\label{Sec:AppendixEigenspectrum}

\subsection{No spin-orbit coupling corrections}
Inserting the ansatz from Eq.~(\ref{SpinPolarized_states}) in the Schr\"{o}dinger equation given by the Hamiltonian~(\ref{Total_Hamiltonian}) for $\Delta_0=\xi_e=0$ and introducing the transformation
\begin{equation}\label{yDEq_Transformation}
\zeta=\zeta(y)=\sqrt{2}\left(y-l_{\mathsmaller{B}}^2k\right)/l_{\mathsmaller{B}}
\end{equation}
yields a system of two differential equations to determine the eigenenergy $E=\epsilon_{ns}(k)$ and eigenstates with given spin and momentum quantum numbers $s$ and $k$. If the $y$-components in Eq.~(\ref{SpinPolarized_states}) are defined as $f^s_{nk}(y)\equiv\tilde{f}_s[\zeta(y)]$ and $g^s_{nk}(y)\equiv\tilde{g}_s[\zeta(y)]$, the system is given by
\begin{equation}\label{yDEq_SpinPolarized}
\begin{aligned}
0 & =\left[\mathcal{C}-E-\frac{2\mathcal{D}}{l_{\mathsmaller{B}}^2}\left(\frac{\zeta^2}{4}-\partial_\zeta^2\right)\right]\left(\begin{array}{c}\tilde{f}_s(\zeta)\\
\tilde{g}_s(\zeta)\end{array}\right)\\
& +\left[\mathcal{M}-\frac{2\mathcal{B}}{l_{\mathsmaller{B}}^2}\left(\frac{\zeta^2}{4}-\partial_\zeta^2\right)\right]\left(\begin{array}{c}\tilde{f}_s(\zeta)\\ -\tilde{g}_s(\zeta)\end{array}\right)\\
& -\frac{\sqrt{2}\mathcal{A}}{l_{\mathsmaller{B}}}\left(\begin{array}{c}\left(s\frac{\zeta}{2}-\partial_\zeta\right)\tilde{g}_s(\zeta)\\ \left(s\frac{\zeta}{2}+\partial_\zeta\right)\tilde{f}_s(\zeta)\end{array}\right)+s\frac{\mu_{\mathsmaller{B}}B}{2}\left(\begin{array}{c}g_{\mathsmaller{\mathrm{e}}}\tilde{f}_s(\zeta)\\ g_{\mathsmaller{\mathrm{h}}}\tilde{g}_s(\zeta)\end{array}\right).
\end{aligned}
\end{equation}
Next, we impose boundary conditions along the $y$-direction, namely
\begin{equation}\label{BC_Bulk_SpinPolarized}
\lim\limits_{\zeta\to\pm\infty}\tilde{f}_s(\zeta)=\lim\limits_{\zeta\to\pm\infty}\tilde{g}_s(\zeta)=0
\end{equation}
for the infinite bulk geometry~(i) and
\begin{equation}\label{BC_FiniteStrip_SpinPolarized}
\tilde{f}_s[\zeta(\pm w/2)]=\tilde{g}_s[\zeta(\pm w/2)]=0
\end{equation}
for the finite-strip geometry~(ii).

We use a finite-difference scheme to solve the system of differential Eqs.~(\ref{yDEq_SpinPolarized}) numerically. In addition, we also apply a second method to obtain and check the solutions of Eq.~(\ref{yDEq_SpinPolarized}): As detailed in Ref.~\onlinecite{Scharf2012:PRB2} and similar to the procedure in Ref.~\onlinecite{Grigoryan2009:PRB}, the general solution of Eq.~(\ref{yDEq_SpinPolarized}) can be written down in the form of parabolic cylindrical functions. By invoking the appropriate boundary conditions on this general solution, one can then obtain a transcendental equation to determine the eigenspectrum and eigenstates. The boundary conditions for a bulk system given by Eq.~(\ref{BC_Bulk_SpinPolarized}), for example, lead to the LLs in Eqs.~(\ref{0LLup})-(\ref{LLdown}) and their corresponding eigenstates, where the parabolic cylindrical functions are reduced to Hermite polynomials.\cite{Scharf2012:PRB2}

\subsection{Spin-orbit coupling corrections}
If SOC corrections $\Delta_0$ and/or $\xi_e$ are considered and the ansatz from Eq.~(\ref{SpinUnpolarized_states}) as well as the transformation in Eq.~(\ref{yDEq_Transformation}) are used, we redefine the functions $f^1_{nk}(y)\equiv\tilde{f}_1[\zeta(y)]$, $g^1_{nk}(y)\equiv\tilde{g}_1[\zeta(y)]$, $f^2_{nk}(y)\equiv\tilde{f}_2[\zeta(y)]$, and $g^2_{nk}(y)\equiv\tilde{g}_2[\zeta(y)]$. Then, the eigenenergy $E=\epsilon_{n}(k)$ of an eigenstate with momentum $k$ is determined by the system of four differential equations
\begin{equation}\label{yDEq_SpinUnpolarized}
\begin{aligned}
0 & =\left[\mathcal{C}-E-\frac{2\mathcal{D}}{l_{\mathsmaller{B}}^2}\left(\frac{\zeta^2}{4}-\partial_\zeta^2\right)\right]\left(\begin{array}{c}\tilde{f}_1(\zeta)\\ \tilde{g}_1(\zeta)\\ \tilde{f}_2(\zeta)\\ \tilde{g}_2(\zeta)\end{array}\right)\\
& +\left[\mathcal{M}-\frac{2\mathcal{B}}{l_{\mathsmaller{B}}^2}\left(\frac{\zeta^2}{4}-\partial_\zeta^2\right)\right]\left(\begin{array}{c}\tilde{f}_1(\zeta)\\ -\tilde{g}_1(\zeta)\\ \tilde{f}_2(\zeta)\\ -\tilde{g}_2(\zeta)\end{array}\right)\\
& -\frac{\sqrt{2}\mathcal{A}}{l_{\mathsmaller{B}}}\left(\begin{array}{c}\left(\frac{\zeta}{2}-\partial_\zeta\right)\tilde{g}_1(\zeta)\\ \left(\frac{\zeta}{2}+\partial_\zeta\right)\tilde{f}_1(\zeta)\\ -\left(\frac{\zeta}{2}+\partial_\zeta\right)\tilde{g}_2(\zeta)\\ -\left(\frac{\zeta}{2}-\partial_\zeta\right)\tilde{f}_2(\zeta)\end{array}\right)+\frac{\mu_{\mathsmaller{B}}B}{2}\left(\begin{array}{c}g_{\mathsmaller{\mathrm{e}}}\tilde{f}_1(\zeta)\\ g_{\mathsmaller{\mathrm{h}}}\tilde{g}_1(\zeta)\\ -g_{\mathsmaller{\mathrm{e}}}\tilde{f}_2(\zeta)\\ -g_{\mathsmaller{\mathrm{h}}}\tilde{g}_2(\zeta)\end{array}\right)\\
& +\Delta_0\left(\begin{array}{c}-\tilde{g}_2(\zeta)\\ \tilde{f}_2(\zeta)\\ \tilde{g}_1(\zeta)\\ -\tilde{f}_1(\zeta)\end{array}\right)-\frac{\i\sqrt{2}\xi_\mathrm{e}}{l_{\mathsmaller{B}}}\left(\begin{array}{c} \left(\frac{\zeta}{2}+\partial_\zeta\right)\tilde{f}_2(\zeta)\\ 0\\ -\left(\frac{\zeta}{2}-\partial_\zeta\right)\tilde{f}_1(\zeta)\\ 0\end{array}\right)
\end{aligned}
\end{equation}
and the boundary conditions Eqs.~(\ref{BC_Bulk_SpinPolarized}) or Eqs.~(\ref{BC_FiniteStrip_SpinPolarized}), where $s$ is to be replaced by the indices 1 and 2. The system of differential Eqs.~(\ref{yDEq_SpinUnpolarized}) is then solved numerically with a finite-difference scheme.

\section{Current operator and dipole matrix elements}\label{Sec:AppendixCurrent}
To calculate the optical conductivity via the Kubo formula given by Eqs.~(\ref{OpticalConductivity}) and~(\ref{CurrentCurrentCorrelationFunction}), we need to know the charge current operator $\hat{\mathbf{I}}$. In the presence of an arbitrary magnetic vector potential $\mathbf{A}(\mathbf{r})$, Eqs.~(\ref{SpinPolarized_Hamiltonian}) and~(\ref{SOC_Hamiltonian}) describing the HgTe QW have to be generalized by
\begin{equation}\label{App2_SpinPolarized_Hamiltonian}
\begin{aligned}
\hat{H}_0=&\mathcal{C}\mathbf{1}_4+\mathcal{M}\Gamma_5-\frac{\mathcal{D}\mathbf{1}_4+\mathcal{B}\Gamma_5}{\hbar^2}\left(\hat{\pi}_x^2+\hat{\pi}_y^2\right)\\
&+\frac{\mathcal{A}\Gamma_1}{\hbar}\hat{\pi}_x+\frac{\mathcal{A}\Gamma_2}{\hbar}\hat{\pi}_y+\frac{\mu_{\mathsmaller{B}}\left[\mathbf{\nabla}\times\mathbf{A}(\mathbf{r})\right]\mathbf{\Gamma}_g}{2},
\end{aligned}
\end{equation}
and
\begin{equation}\label{App2_SOC_Hamiltonian}
\hat{H}_\mathrm{SOC}=\Delta_0\Gamma_\mathrm{BIA}+\frac{\xi_e}{\hbar}\left(\Gamma_\mathrm{SIA1}\hat{\pi}_x+\Gamma_\mathrm{SIA2}\hat{\pi}_y\right),
\end{equation}
respectively. Here, the kinetic momentum operator $\hat{\bm{\pi}}=\hat{\mathbf{p}}+e\mathbf{A}(\mathbf{r})$, $\mathbf{\Gamma}_g=\left(\Gamma_g^x,\Gamma_g^y,\Gamma_g^z\right)$ and the additional $4\times4$ matrices
\begin{equation}\label{App_Matrices}
\begin{aligned}
\Gamma^x_g=\left(\begin{array}{cc}
         0 & g_{\mathsmaller{\parallel}}\mathbf{1}_2 \\
         g_{\mathsmaller{\parallel}}\mathbf{1}_2 & 0 \\
         \end{array}\right),
\Gamma^y_g=\left(\begin{array}{cc}
         0 & -\i g_{\mathsmaller{\parallel}}\mathbf{1}_2 \\
         \i g_{\mathsmaller{\parallel}}\mathbf{1}_2 & 0 \\
         \end{array}\right),
\end{aligned}
\end{equation}
as well as the effective in-plane g-factor $g_{\mathsmaller{\parallel}}$ have been introduced.\cite{Koenig2008:JPSJ}

For an arbitrary (normalized) state $\Psi(\mathbf{r})$, the corresponding energy expectation value of the total Hamiltonian $\hat{H}$, given by Eq.~(\ref{Total_Hamiltonian}) and generalized by Eqs.~(\ref{App2_SpinPolarized_Hamiltonian}) and~(\ref{App2_SOC_Hamiltonian}), as a functional of the vector potential $\mathbf{A}(\mathbf{r})$ can be obtained as
\begin{equation}\label{App2_EnergyFunctional}
E\left[\mathbf{A}\right]=\sum\limits_{\alpha\beta}\int\d^2r\;\Psi_\alpha^*(\mathbf{r})H_{\alpha\beta}\Psi_\beta(\mathbf{r}),
\end{equation}
where the sums over $\alpha$ and $\beta$ refer to the four bands considered, that is, $\ket{E\uparrow}$, $\ket{H\uparrow}$, $\ket{E\downarrow}$, $\ket{H\downarrow}$. The particle current density $\mathbf{j}(\mathbf{r})$ of this state $\Psi(\mathbf{r})$ can be determined by a variational method:
\begin{equation}\label{App2_Variation}
\delta E=E\left[\mathbf{A}+\delta\mathbf{A}\right]-E\left[\mathbf{A}\right]=e\int\d^2r\;\mathbf{j}(\mathbf{r})\delta\mathbf{A}(\mathbf{r}).
\end{equation}
This procedure yields the probability current density $\mathbf{j}(\mathbf{r})=\mathbf{j}_\mathrm{e}(\mathbf{r})+\mathbf{j}_\mathrm{i}(\mathbf{r})$ composed of the external current density,
\begin{equation}\label{App2_ExternalCurrent}
\begin{array}{l}
\mathbf{j}_\mathrm{e}(\mathbf{r})=\\
\sum\limits_{\alpha\beta}\biggl\{\frac{\i}{\hbar}\left[\mathcal{D}\left(\mathbf{1}_4\right)_{\alpha\beta}+\mathcal{B}\left(\Gamma_5\right)_{\alpha\beta}\right]\left[\Psi_\alpha^*\left(\mathbf{\nabla}\Psi_\beta\right)-\left(\mathbb{\nabla}\Psi_\alpha^*\right)\Psi_\beta\right]\\
+\frac{\mathcal{A}}{\hbar}\left(\mathbf{\Gamma}\right)_{\alpha\beta}\Psi_\alpha^*\Psi_\beta-\frac{2e}{\hbar^2}\left[\mathcal{D}\left(\mathbf{1}_4\right)_{\alpha\beta}+\mathcal{B}\left(\Gamma_5\right)_{\alpha\beta}\right]\Psi_\alpha^*\Psi_\beta \mathbf{A}\\
+\frac{\xi_e}{\hbar}\left(\mathbf{\Gamma}_\mathrm{SIA}\right)_{\alpha\beta}\Psi_\alpha^*\Psi_\beta\biggr\},
\end{array}
\end{equation}
where $\mathbf{\Gamma}=\left(\Gamma_1,\Gamma_2,0\right)$ and $\mathbf{\Gamma}_\mathrm{SIA}=\left(\Gamma_\mathrm{SIA1},\Gamma_\mathrm{SIA2},0\right)$, and the internal current density,
\begin{equation}\label{App2_InternalCurrent}
\mathbf{j}_\mathrm{i}(\mathbf{r})=\frac{\mu_{\mathsmaller{B}}}{2e}\mathbf{\nabla}\times\left[\sum\limits_{\alpha\beta}\Psi_\alpha^*\left(\mathbf{\Gamma}_g\right)_{\alpha\beta}\Psi_\beta\right].
\end{equation}
As we are dealing with a 2D system, Eqs.~(\ref{App2_ExternalCurrent}) and~(\ref{App2_InternalCurrent}) are to be read as applying only to the $x$- and $y$-components.

We note that the external current density given by Eq.~(\ref{App2_ExternalCurrent}) could also have been obtained by calculating the velocity operator $\hat{\mathbf{v}}=\left[\hat{\mathbf{r}},\hat{H}\right]/\i\hbar$ and using $\mathbf{j}=\left[\Psi^*\left(\hat{\mathbf{v}}\Psi\right)+\left(\hat{\mathbf{v}}\Psi\right)^*\Psi\right]/2$. Since for the parameters and magnetic fields investigated in this work, the internal current is very small compared to the external current, we neglect its contribution and only consider the external current density from now on.

By promoting the wave functions $\Psi_\alpha^*(\mathbf{r})$ and $\Psi_\beta(\mathbf{r})$ in Eq.~(\ref{App2_ExternalCurrent}) to field operators $\hat{\Psi}_\alpha^\dagger(\mathbf{r})$ and $\hat{\Psi}_\beta(\mathbf{r})$, we obtain the current density operator $\hat{\mathbf{j}}_\mathrm{e}(\mathbf{r})$. As basis for the field operators, we use the eigenstates of the respective system, that is, the states given by Eq.~(\ref{SpinPolarized_states}) and determined from Eq.~(\ref{yDEq_SpinPolarized}) for a system where spin is a good quantum number and the states given by Eqs.~(\ref{SpinUnpolarized_states}) and~(\ref{yDEq_SpinUnpolarized}) if SOC corrections are taken into account. The charge current operator $\hat{\mathbf{I}}$ is then obtained from
\begin{equation}\label{ChargeCurrent_General}
\hat{\mathbf{I}}=-e\int\d^2r\hat{\mathbf{j}}_\mathrm{e}(\mathbf{r}),
\end{equation}
which yields the components
\begin{equation}\label{ChargeCurrent_SpinPolarized}
\hat{I}_l=-e\sum\limits_{k,s,n,n'}d_{nn'}^{l,s}(0,k)\hat{c}^\dagger_{nks}\hat{c}_{n'ks}
\end{equation}
and
\begin{equation}\label{ChargeCurrent_SpinUnpolarized}
\hat{I}_l=-e\sum\limits_{k,n,n'}d_{nn'}^{l}(0,k)\hat{c}^\dagger_{nk}\hat{c}_{n'k}
\end{equation}
in the $l$-direction for the case with and without SOC corrections. Here, the operators $\hat{c}^\dagger_{nks}$ ($\hat{c}_{nks}$) and $\hat{c}^\dagger_{nk}$ ($\hat{c}_{nk}$) create (destroy) an electron in a state given by Eqs.~(\ref{SpinPolarized_states}) and ~(\ref{SpinUnpolarized_states}), respectively. The corresponding dipole matrix element $d_{nn'}^{l,s}(0,k)$ is calculated by replacing $\Psi_\alpha^*(\mathbf{r})$ and $\Psi_\beta(\mathbf{r})$ in Eq.~(\ref{App2_ExternalCurrent}) with the eigenstates $\left[\Psi^s_{nk}(\mathbf{r})\right]_\alpha^*$ and $\left[\Psi^s_{n'k}(\mathbf{r})\right]_\beta$ from Eq.~(\ref{SpinPolarized_states}) and integrating over $\d^2r$. In the case of an infinite bulk system, analytical formulas can be derived for the dipole matrix elements using the eigenstates given in Ref.~\onlinecite{Scharf2012:PRB2}. However, those dipole matrix elements are quite cumbersome and not particularly elucidating. If SOC corrections are considered, the dipole matrix element $d_{nn'}^{l}(0,k)$ is determined in a similar way from the eigenstates $\left[\Psi_{nk}(\mathbf{r})\right]_\alpha^*$ and $\left[\Psi_{n'k}(\mathbf{r})\right]_\beta$ of Eq.~(\ref{SpinUnpolarized_states}).

\bibliography{BibTopInsAndTopSup}

\end{document}